\documentclass[doublecol]{epl2} 

\usepackage{color}
\usepackage{fdsymbol}
\usepackage{graphicx}
\usepackage{dcolumn}
\usepackage{bm}

\title{Highly Clustered Complex Networks in the Configuration Model: \\Random Regular Small-World Network}
\shorttitle{Highly Clustered Complex Networks in the Configuration Model}

\author{Wonhee Jeong \and Hoseung Jang \and Unjong Yu\thanks{E-mail: \email{uyu@gist.ac.kr}}}

\institute{
  Department of Physics and Photon Science, Gwangju Institute of Science and Technology, Gwangju 61005, South Korea
}

\abstract{
We propose a method to make a highly clustered complex network within the configuration model. Using this method, we generated highly clustered random regular networks and analyzed the properties of them. We show that highly clustered random regular networks with appropriate parameters satisfy all the conditions of the small-world network: connectedness, high clustering coefficient, and small-world effect. We also study how clustering affects the percolation threshold in random regular networks. In addition, the prisoner's dilemma game is studied and the effects of clustering and degree heterogeneity on the cooperation level are discussed.
}

\begin{document}

\maketitle

\section{\label{sec:level1}Introduction}

Recently, complex networks have played important roles in various fields~\cite{Strogatz2001,biamonte2019complex,lynn2019physics,fortunato2018science,boccaletti2014structure}. There are, especially, a lot of researches on the various types of real-world network structures \cite{Strogatz2001,fortunato2018science,boccaletti2014structure}. As a result, studies to generate network models that resemble various kinds of real-world networks attract a lot of attention. Many methods were proposed to generate complex networks: rewiring model, 
growing model, configuration model, etc~\cite{watts1998collective,PhysRevE.72.056128,Barabasi509,PhysRevE.65.026107,PhysRevE.65.057102,4150252,4624849,PhysRevE.95.052303,PhysRevE.89.062814,PhysRevLett.96.040601,PhysRevLett.103.058701,RevModPhys.80.1275,boccaletti2006complex}.

The Watts-Strogatz (WS) model generates a small-world network based on the rewiring method~\cite{watts1998collective}. The small-world network has a high clustering coefficient and short average path length, but it has restriction in the degree distribution~\cite{watts1998collective,PhysRevE.72.056128}.

The Barab\'{a}si-Albert (BA) model generates a scale-free network using the growing model with the preferential attachment~\cite{Barabasi509}. It produces the power-law degree distribution, which is frequently observed in a real-world network such as citation networks, metabolic networks, and the Internet~\cite{jeong2000large,HAJRA200544,PhysRevE.68.026113,PhysRevLett.87.258701,PhysRevE.66.035103}. However, the network generated by the BA model has a vanishing clustering coefficient in contrast with real-world networks. To overcome this problem of the BA model, many researchers proposed modified BA models~\cite{PhysRevE.65.026107,PhysRevE.65.057102,4150252,4624849} that have a high clustering coefficient maintaining the power-law degree distribution.

The configuration model is a method to make a random network for a given arbitrary degree sequence.
This model has the advantage of being able to use the degree sequence of the real-world network. However,
this model also has the problem of vanishing clustering coefficient like the BA model. Although this model has also been studied as much as the BA model~\cite{PhysRevE.95.052303,PhysRevE.89.062814,PhysRevLett.96.040601}, the method to control clustering in the configuration model is rare in contrast to the BA model. Newman proposed a method to generate a highly clustered network model from two sequences for the degree and the number of corners of triangles by modifying the configuration model~\cite{PhysRevLett.103.058701}, but generally the distribution of the number of corners of triangles is not known in advance. Therefore, in this paper, we propose a highly clustered configuration model that requires only the degree sequence.

Using this model, we made highly clustered random regular networks, which are special cases of the configuration model~\cite{bollobas2001random,Bollobas80,Steger99}. The highly clustered random regular network has a wide range of clustering coefficient, which can be estimated in a certain range. This network also has a giant component and a short average path length. We show that it is the random regular small-world network.

In the network generated by our model, we calculated a percolation threshold of the site percolation. We show that the percolation threshold increases with increasing clustering coefficient. We also studied the prisoner's dilemma game in four networks (BA network, highly clustered BA network, random regular network, and highly clustered random regular network) and compared the cooperation level for each network.

\section{\label{sec:level2}Model of highly clustered random regular network}

\begin{figure}
\centering
\includegraphics[angle=0,width=1\columnwidth]{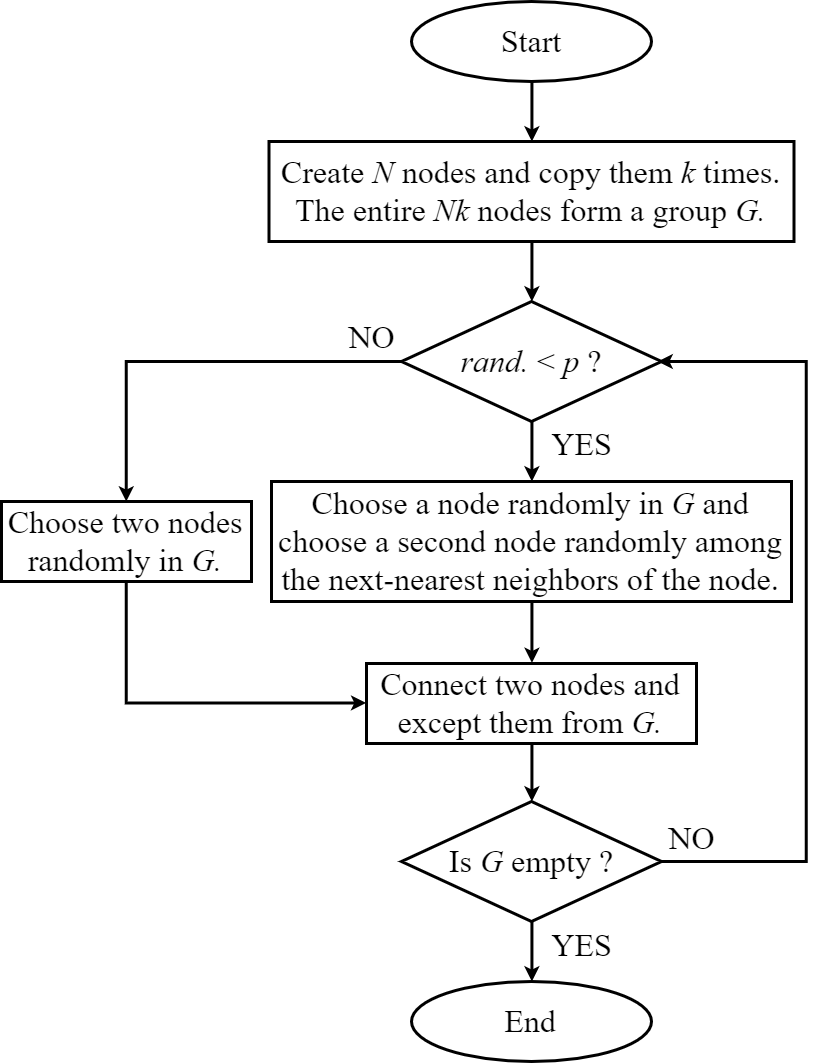}
\caption{The algorithm diagram of the model of highly clustered random regular network.}\label{diagram}
\end{figure}

We propose a method to make a highly clustered random $k$-regular network with $N$ nodes by modifying the configuration model for a random regular network~\cite{bollobas2001random,Bollobas80,Steger99}. We assume that $k \geqq 3$ and $Nk$ is even~\cite{bollobas2001random,Bollobas80,Steger99}. The algorithm is as follows, \\ \\
(1) Make $N$ nodes.\\
(2) Copy $N$ nodes by $k$. These $Nk$ nodes are a group $G$.\\
(3) Make a triangle (T) with probability $p$ and make a random pair (R) with probability ($1-p$).\\
\indent (T-1) Choose a node randomly in $G$.\\
\indent (T-2) Choose a second node randomly among the next nearest neighbors of the node.\\
\indent (T-3) Connect the two nodes and except them from $G$. \\
\indent (T-4) If the triangle connection is impossible, make a random pair.\\
\indent (R-1) Choose two nodes randomly in $G$.\\
\indent (R-2) Connect the two nodes and except them from $G$.\\
(4) Repeat step (3) until the group $G$ is empty.\\ \\
See also the algorithm diagram in Fig.~\ref{diagram}. Note that the self-loop and multiple connections are not allowed. When $p=0$, our model reduces to the conventional random $k$-regular network. It is worth mentioning that this method can be applied to any degree sequence in general.

\section{\label{sec:level3}Properties of the network}

\begin{figure*}[tb]
\centering
\includegraphics[angle=270,width=2\columnwidth]{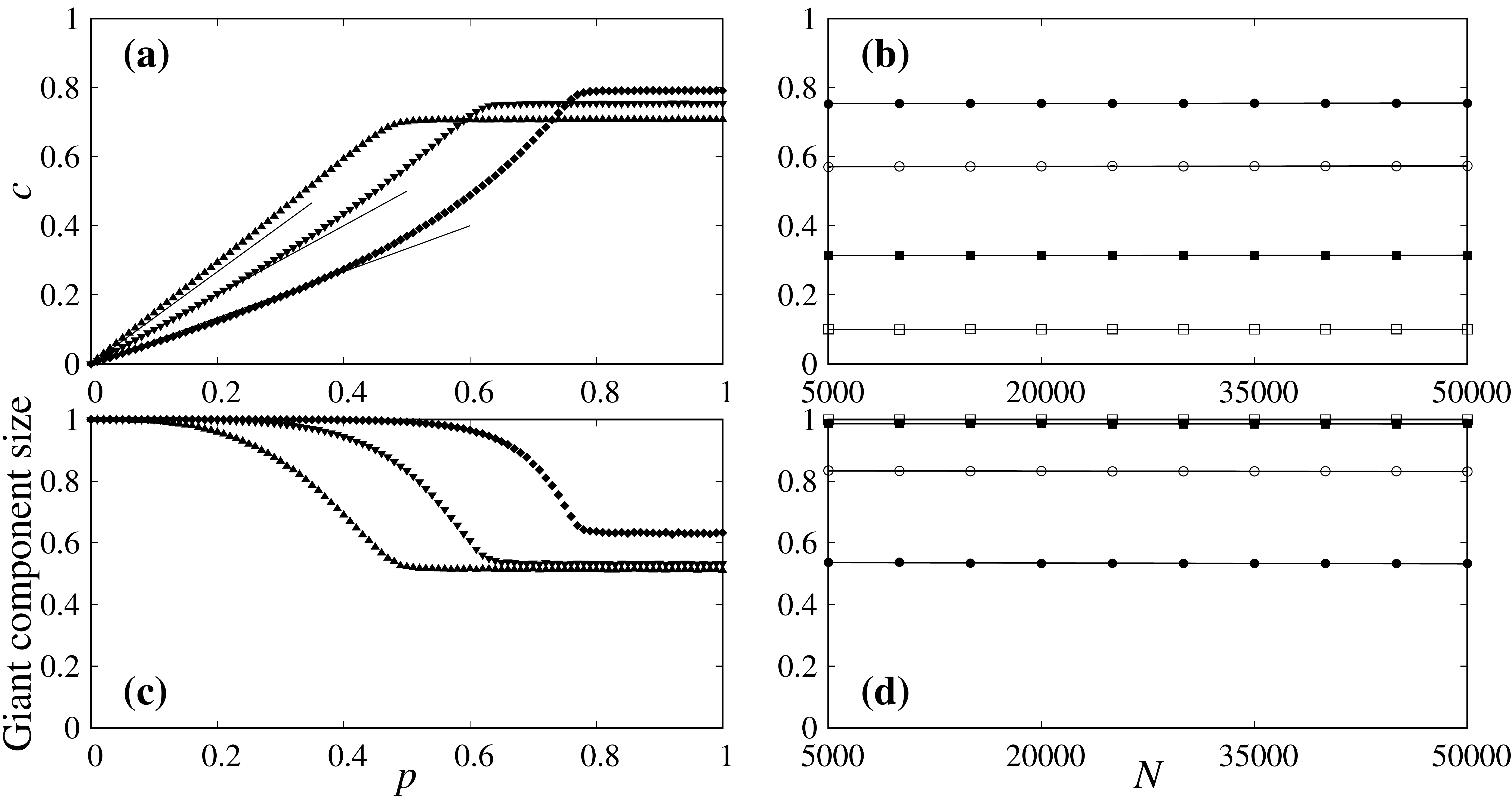}
\caption{(a) The clustering coefficient as a function of $p$ for $k=3$ ($\medblacktriangleup$), $k=4$ ($\medblacktriangledown$), and $k=6$ ($\medblackdiamond$). The solid lines are from Eq.~(\protect\ref{EQ5}). 
(b) The clustering coefficient as a function of $N$ for $p=0.7$ ($\medblackcircle$), $p=0.5$ ($\medcircle$), $p=0.3$ ($\medblacksquare$), and $p=0.1$ ($\medsquare$). The solid lines are guide to the eyes. 
(c) The giant component size as a function of $p$ for $k=3$ ($\medblacktriangleup$), $k=4$ ($\medblacktriangledown$), and $k=6$ ($\medblackdiamond$). 
(d) The giant component size as a function of $N$ for $p=0.7$ ($\medblackcircle$), $p=0.5$ ($\medcircle$), $p=0.3$ ($\medblacksquare$), and $p=0.1$ ($\medsquare$). The solid lines are guide to the eyes.
The number of total nodes $N$ is $10\,000$ in (a) and (b). The degree $k$ is 4 in (b) and (d).
Each point was averaged over $100$ independently generated networks.
}
\label{fig1}
\end{figure*}

Complex networks have a huge number of nodes and edges, and their network topologies are nontrivial. Therefore, many kinds of parameters have been proposed to characterize complex networks~\cite{erdos1959random,freeman1977set,watts1998collective,PhysRevLett.89.208701}. Among them, we introduce three parameters and study them for the network made by the algorithm proposed in the previous section.

The first one is the clustering coefficient $c$, which indicates the cliquishness of the neighbors of a node~\cite{RevModPhys.80.1275,watts1998collective}. Most of the social networks have a considerable value of $c$~\cite{RevModPhys.74.47}. The clustering coefficient $c$ is defined by 
\begin{eqnarray}
c = \frac{1}{N} \sum_{i} c_i ~\mbox{ with } c_i = \frac{e_i}{k_i (k_i-1) / 2} , \label{EQ2}
\end{eqnarray}
where $N$ is the number of total nodes in the network, $k_i$ is the degree of node $i$, and $e_i$ is the number of edges between neighbors of node $i$~\cite{RevModPhys.74.47,watts1998collective,RevModPhys.80.1275}.

We calculated the clustering coefficient as a function of $p$ and $N$, as shown in Figs.~\ref{fig1}(a) and \ref{fig1}(b). The clustering coefficient increases monotonically as $p$ increases, but it is saturated at a specific value of $p$. The point where $c$ is saturated depends on the number of neighbors $k$. The clustering coefficient can be estimated assuming that $p$ is sufficiently small such that no cliques are formed. The probability $P_e$ that the edge exists between two neighbors of the node is given approximately by
\begin{eqnarray}
P_e = 2p + 2(1-p)\left(\frac{k-1}{N-2}\right) . \label{EQ3}
\end{eqnarray}
The first and last terms on the right-hand side are from the triangle connection and random pair connection, respectively; by the triangle connection the edge is made between the neighbors almost always, and 
the probability becomes $(k-1)/(N-2)$ for the random pair connection.
The last term can be ignored in most cases of $N \gg k$. That is, $P_e \approx 2p$.
The factor of 2 represents two chances to be connected from two nodes connected by the edge.
Since the first edge connection should be by the random pair connection, the number of possible edges should be multiplied by $(k-1)$. Therefore, by Eq.~(\ref{EQ2}), the clustering coefficient $c$ is 
\begin{eqnarray}
c = \frac{(k-1) P_e}{k(k-1)/2} 
  \approx \frac{4p}{k} . \label{EQ5}
\end{eqnarray}
The solid lines in Fig.~\ref{fig1}(a) represent this equation for each $k$, which are consistent with numerical results for small $p$. When $p$ is large, there appear deviations from Eq.~(\ref{EQ5}) as expected.

The second property is the existence and size of the giant component. The giant component is the largest component of the network whose size has a finite fraction of the network irrespective of the number of total nodes of the network.

The giant component size of the network is shown as a function of $p$ in Fig.~\ref{fig1}(c). The network is connected for small $p$, the giant component size decreases as $p$ increases, and then it is saturated at a specific $p$. Since our model network tends to have more cliques when $p$ is large, the network is more likely to be disconnected and the giant component size decreases.  Comparison of Figs.~\ref{fig1}(a) and \ref{fig1}(c) shows that both the clustering coefficient and the giant component size are saturated at the same value of $p$ for each $k$. Figures~\ref{fig1}(b) and \ref{fig1}(d) show that the clustering coefficient and the giant component size are constant with the network size $N$, which confirms that the network generated by our model is consistently clustered and always has the giant component.

The last parameter is the average path length (characteristic path length) $l$. This is defined by
\begin{eqnarray}
l = \frac{1}{N(N-1)} \sum_{i \neq j} L(i,j) ,\label{EQ6}
\end{eqnarray}
where $L(i,j)$ indicates the shortest path length between nodes $i$ and $j$. 
If a complex network follows the relation $l \sim \ln N$, it is called as the small-world effect.
The Erd\H{o}s-R\'{e}nyi random network and random regular (RR) networks were shown to have this property for large $N$~\cite{bollobas2001random,1354658,PhysRevE.93.062309}.

\begin{figure}[tb]
\centering
\includegraphics[angle=270,width=1\columnwidth]{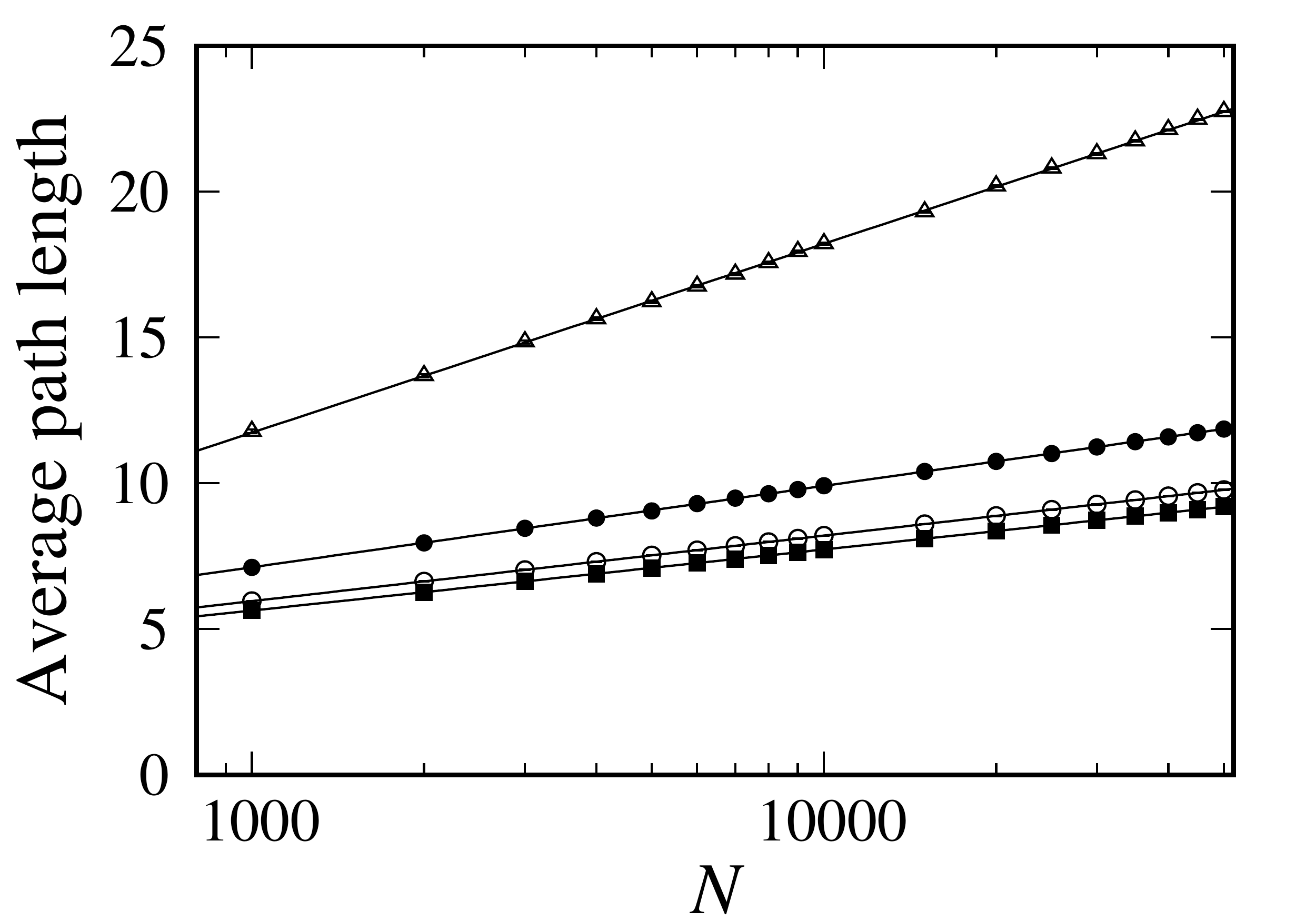}
\caption{The average path length $l$ as a function of the number of total nodes $N$. Values of $p$ are $0.7$, $0.3$, $0.1$, and $0.0$ from top to bottom. The solid lines are from the fitting of $l \sim \ln N$. The degree $k$ is $4$. Each point was averaged over $100$ independently generated networks.}\label{fig2}
\end{figure}

We measured the average path length $l$ of the highly clustered random regular network. Since an unconnected network has infinite average path length, we measured the average path length within the giant component when the network is unconnected. Figure~\ref{fig2} shows that the average path length follows the relation $l \sim \ln N$ very well regardless of $p$.

In general a network is called the small-world network if (i) the network is connected, (ii) $l \sim \ln N$, and (iii) $c$ is high~\cite{watts1998collective,newmanch2,PhysRevLett.84.3201,NEWMAN1999341}. Therefore, we conclude that the network made by the algorithm of the previous section with appropriate $p$ is the random regular small-world network,  based on the results of Figs.~\ref{fig1} and \ref{fig2}. Although a random regular small-world network is rare in real-world, it can be used to study the effects of clustering and degree heterogeneity in complex networks.

\section{\label{sec:level4}Percolation}

\begin{figure}[tb]
\centering
\includegraphics[angle=270,width=1\columnwidth]{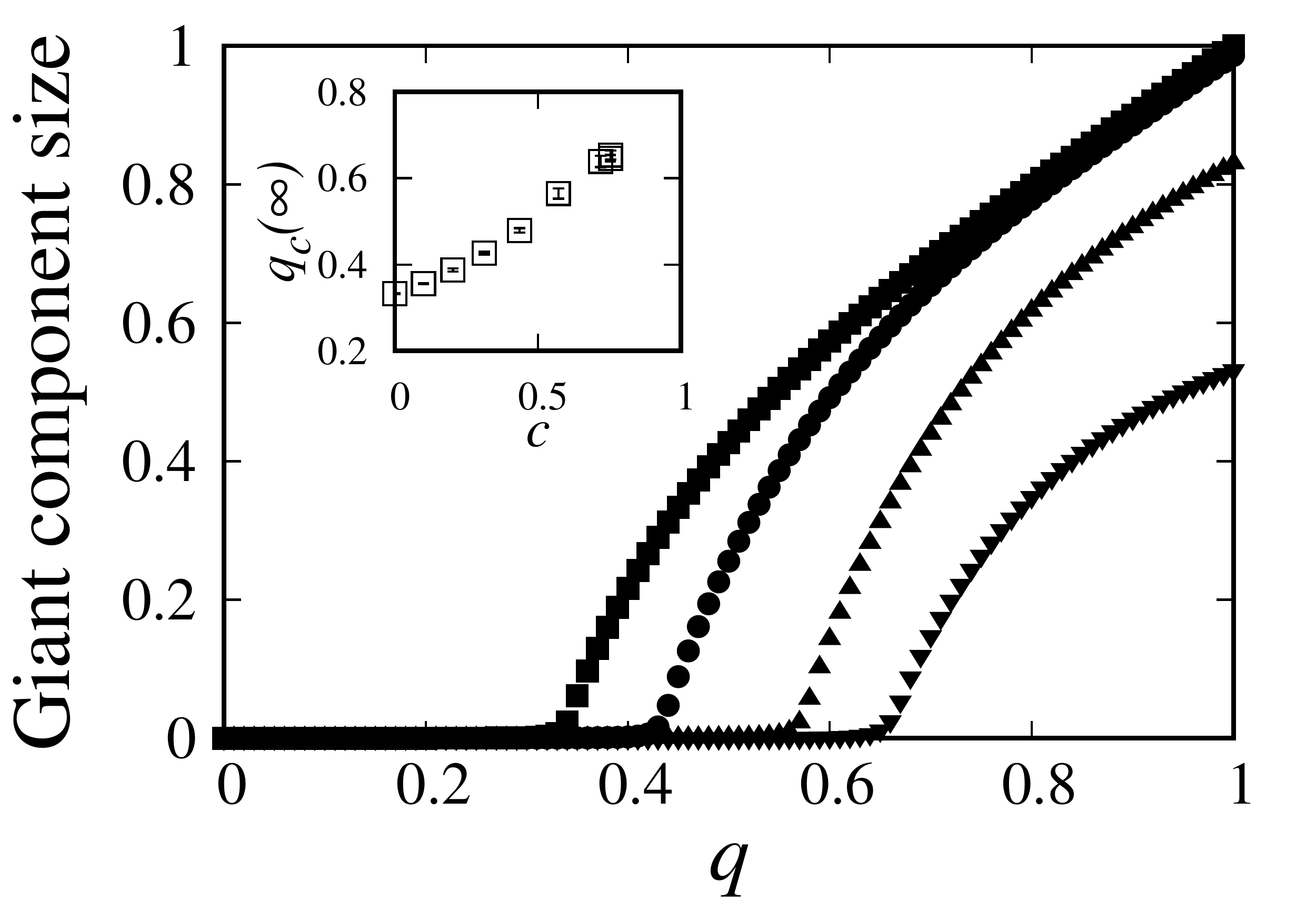}
\caption{The giant component size as a function of the occupation probability $q$ in RR networks with and without clustering. The values of $p$ for square, circle, triangle, and inverse triangle are $0.0$, $0.3$, $0.5$, and $0.7$, respectively. The number of total nodes $N$ is $250\,000$ and  the degree $k$ is $4$. (inset) Percolation threshold $q_{c}(\infty)$ as a function of the clustering coefficient $c$.}\label{fig3}
\end{figure}

Percolation is one of the most important subjects in physics and network science~\cite{stauffer2014introduction,Callaway00,Cohen00}. Among the many types of percolation models, we consider the site percolation: a site (node) is occupied with probability $q$ and empty with probability ($1-q$)~\cite{stauffer2014introduction}. We study the site percolation model in the highly clustered random regular network generated by our model using the Newman-Ziff algorithm~\cite{PhysRevLett.85.4104,PhysRevE.64.016706}. The Newman-Ziff algorithm uses a convolution with the binomial distribution,
\begin{eqnarray}
\langle Q(q) \rangle = \sum_{n} \frac{N!}{n! (N-n)!} q^{n} (1-q)^{N-n} \langle Q(n) \rangle, \label{EQ8}
\end{eqnarray}
where $Q(q)$ is any physical quantity for a given $q$. Variable $n$ represents the number of occupied sites in the network ($0 \leq n \leq N$). Since $Q(q)$ is derived by $Q(n)$, we can get $Q(q)$ and $dQ(q)/dq$ for an arbitrary $q$ when we obtain $Q(n)$ once~\cite{choi2019newman}. Using Eq.~(\ref{EQ8}), we calculated the giant component size $P_\infty(q)$ for a given $q$. The value of $q$ where the giant component size begins to have nonzero value in infinite network is called the percolation threshold $q_{c}(\infty)$, which can be found by the finite-size scaling,
\begin{eqnarray}
\left[ q_{c}(N) - q_{c}(\infty) \right] \sim \frac{1}{N^{a}}. \label{EQ9}
\end{eqnarray}
Estimate of the percolation threshold in finite network $q_{c}(N)$ is obtained by the value of $q$ that gives the maximum of $dP_\infty(q)/dq$ in the network with $N$ nodes, and $a$ is a fitting parameter ($a\geqq0$)~\cite{stauffer2014introduction,PhysRevLett.85.4104,PhysRevE.64.016706}. 

The giant component size $P_\infty(q)$ presented in Fig.~\ref{fig3} shows that the networks exhibit continuous percolation transitions. The percolation threshold obtained by Eq.~(\ref{EQ9}) is presented in the inset of Fig.~\ref{fig3}. We confirm that the random $k$-regular network without clustering ($p=0$) has $q_{c}(\infty)=1/(k-1)$ within error bars consistently with an analytic result~\cite{Cohen00}. The percolation threshold $q_{c}(\infty)$ increases as clustering coefficient $c$ increases. Though not shown here, we observed a similar increase of $q_{c}(\infty)$ with clustering also for the bond percolation model.

\section{\label{sec:level5}Prisoner's dilemma game}

\begin{figure}[tb]
\centering
\includegraphics[angle=270,width=1\columnwidth]{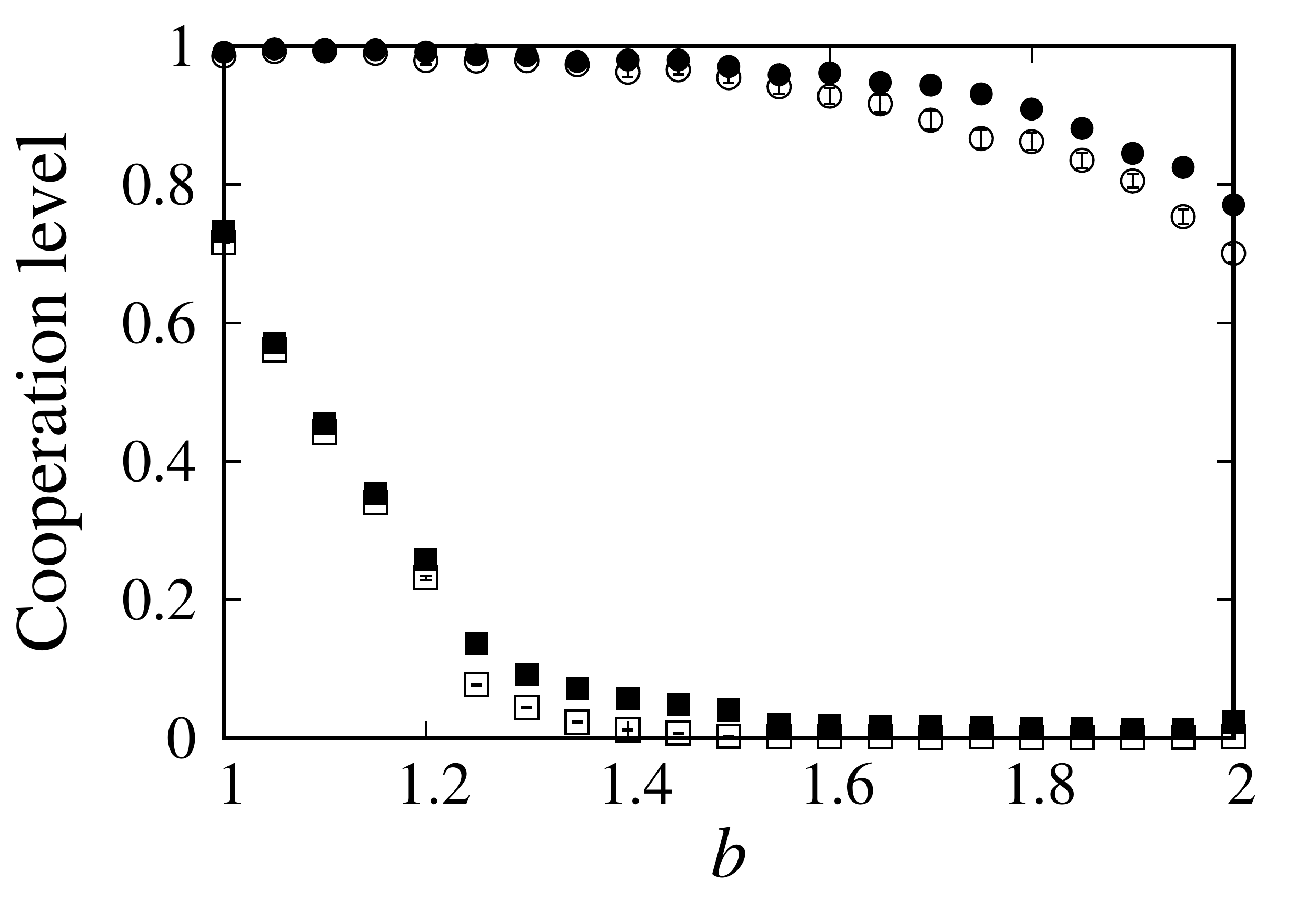}
\caption{The cooperation level as a function of the temptation $b$ in the PD game. The filled circle is for the Holme-Kim network (highly clustered scale-free network), the empty circle is for the BA network, the filled square is for the highly clustered RR network, and the empty square is for the RR network. the number of total nodes (agents) is $10\,000$ and the average degree is $4$. Each point was averaged over $100$ independent networks. The error bars are smaller than the symbol size.}\label{fig4}
\end{figure}

The prisoner's dilemma (PD) game has attracted much attention of many biologists, psychologists, mathematicians, physicists, etc. This game describes the situation where an agent is tempted to choose the defective strategy that is beneficial to itself but detrimental to the whole society.
Especially, the effect of the network topology on the cooperation level is one of the important questions; the cooperation level, in general, depends on the generation method even when the degree distributions are the same~\cite{PhysRevLett.95.098104,CHEN2007379,szabo.am,PhysRevE.79.016106,PhysRevE.73.067103,PhysRevE.72.047107,PhysRevLett.98.108103,Xia_2015,chen2016evolution,chen2015impact,perc2017statistical,li2018punishment,wang2015evolutionary,li2019effect}. In this paper, we compare the cooperation level (proportion of cooperators) in the four kinds of networks: BA network, Holme-Kim network, RR network, and highly clustered RR network (our model)~\cite{Barabasi509,PhysRevE.65.026107,CHEN2007379}. The Holme-Kim network, which has a high clustering coefficient and scale-free degree distribution, is made by modifying the growing method of the BA network~\cite{PhysRevE.65.026107}.

In the PD game on the network, the node and the edge indicate the agent and the relationship between the agents, respectively. Each agent always interacts with the agents connected by edges (neighbors). The PD game simulation begins from a network whose agents choose their strategies between cooperation ($C$) and defection ($D$) uniformly at random. One generation of the simulation is defined as follows. \\ \\
(1) Each agent plays the PD games with each of its neighbors and its payoff is accumulated. \\
(2) Each agent updates its strategy according to the transition probability. \\ \\
A transient time of $20\,000$ generations is enough for the system to reach an equilibrium. After the transition time, $5000$ more generations were simulated and averaged to reduce statistical error. The payoffs of the agent and its neighbor in step (1) are given by the payoff 
\newpage
\noindent
matrix~\cite{nowak1992evolutionary,PhysRevLett.95.098104,CHEN2007379} of
\begin{center}
\begin{tabular}{c|c c}
 & $C$ & $D$ \\
\hline
$C$ & $1$ & $0$ \\
$D$ & $b$ & $0$ \\
\end{tabular} .
\end{center}
This is a weak version of the PD game and the temptation parameter $b$ should satisfy $1 < b < 2$. In step (2), the transition probability  that the agent $i$ imitates the strategy of the agent $j$ is given by
\begin{eqnarray}
 P_{i \rightarrow j} = \begin{cases} \frac{\pi_j-\pi_i}{b \times \mathrm{max}\{k_i,k_j\}} & \text{if}\; \pi_j > \pi_i  \\ 0 & \text{if}\; \pi_j \leq \pi_i \end{cases}, \label{EQ10}
\end{eqnarray}
where $\pi_i$ is the payoff of the agent $i$~\cite{PhysRevLett.95.098104,CHEN2007379}. The agent $j$ is selected randomly among the neighbors of the agent $i$.

Figure~\ref{fig4} shows the cooperation level of the PD game in the four kinds of networks. All of these networks are connected networks.
The parameters of the Holme-Kim network and the highly clustered RR network are adjusted to have similar values of the clustering coefficient to be $c=0.2026(4)$ and $c=0.2034(3)$, respectively. The average number of neighbors is fixed to be $\langle k \rangle = 4$ for all the networks.

As shown in Fig.~\ref{fig4}, the cooperation level decreases as the temptation parameter $b$ increases in all the four networks, as expected. 
The cooperation level of the scale-free network, which has higher degree heterogeneity than the RR network, is much higher. Within the same category of the network (RR or scale-free network), the higher the clustering, the higher the cooperation level.
As the clustering increases, triangles consisting of cooperators are formed more easily, and the cooperators are more likely to survive in the network.
However, the enhancement of the cooperation level by clustering is small. (Since the error bars of the cooperation level in Fig.~\ref{fig4} are very small, the differences in the cooperation level are statistically significant.)
Thus, we conclude that the degree heterogeneity is more influential factor than the clustering in terms of promotion of cooperation.

\section{\label{sec:level6}Conclusion}

We proposed a method to make a highly clustered network within the configuration model. The clustering coefficient can be controlled in a wide range by adjusting the parameter $p$. We generated highly clustered random regular networks by the method and analyzed their properties. 
The clustering coefficient increases with $p$ and it does not depend on the network size at least for $N\geqq5000$. The networks become disconnected for large $p$, but the giant component exists always for the whole range of $0\leqq p \leqq 1$. The average path length $l$ was shown to be proportional to $\ln N$. As a result, we showed that the highly clustered random regular network made in this work is the random regular small-world network for appropriate parameter $p$. We also calculated the percolation threshold of the infinite-size highly clustered random regular network. The percolation threshold increases as the clustering coefficient increases for both the site and bond percolation problems. In addition, we compared the cooperation level of the prisoner's dilemma game in four kinds of networks. We conclude that clustering promotes cooperation among agents, but the influence of clustering on the cooperation is much smaller than that of degree heterogeneity of the network.

\acknowledgments
This work was supported by GIST Research Institute (GRI) grant funded by the GIST in 2019.

\bibliography{references}

\begin{thebibliography}{10}
\expandafter\ifx\csname url\endcsname\relax\def\url#1{\texttt{#1}}\fi

\bibitem{Strogatz2001}
\Name{Strogatz S.~H.} \REVIEW{Nature}{410}{2001}{268}.

\bibitem{biamonte2019complex}
\Name{Biamonte J., Faccin M. \and De~Domenico M.} \REVIEW{Commun.
  Phys.}{2}{2019}{53}.

\bibitem{lynn2019physics}
\Name{Lynn C.~W. \and Bassett D.~S.} \REVIEW{Nat. Rev. Phys.}{1}{2019}{318}.

\bibitem{fortunato2018science}
\Name{Fortunato S., Bergstrom C.~T., B{\"o}rner K., Evans J.~A., Helbing D.,
  Milojevi{\'c} S., Petersen A.~M., Radicchi F., Sinatra R., Uzzi B.,
  Vespignani A., Waltman L., Wang D. \and Barab{\'a}si A.-L.}
  \REVIEW{Science}{359}{2018}{eaao0185}.

\bibitem{boccaletti2014structure}
\Name{Boccaletti S., Bianconi G., Criado R., Del~Genio C.~I.,
  G\'omez-Garde\~nes J., Romance M., Sendi\~na Nadal I., Wang Z. \and Zanin M.}
  \REVIEW{Phys. Rep.}{544}{2014}{1}.

\bibitem{watts1998collective}
\Name{Watts D.~J. \and Strogatz S.~H.} \REVIEW{Nature}{393}{1998}{440}.

\bibitem{PhysRevE.72.056128}
\Name{Santos F.~C., Rodrigues J.~F. \and Pacheco J.~M.} \REVIEW{Phys. Rev.
  E}{72}{2005}{056128}.

\bibitem{Barabasi509}
\Name{Barab\'{a}si A.-L. \and Albert R.} \REVIEW{Science}{286}{1999}{509}.

\bibitem{PhysRevE.65.026107}
\Name{Holme P. \and Kim B.~J.} \REVIEW{Phys. Rev. E}{65}{2002}{026107}.

\bibitem{PhysRevE.65.057102}
\Name{Klemm K. \and Egu\'{\i}luz V.~M.} \REVIEW{Phys. Rev.
  E}{65}{2002}{057102}.

\bibitem{4150252}
\Name{{Fu} P. \and {Liao} K.} \Book{An evolving scale-free network with large
  clustering coefficient} in proc. of \Book{9th International Conference on
  Control, Automation, Robotics and Vision} 2006 pp. 1--4.

\bibitem{4624849}
\Name{{Wang} J. \and {Rong} L.} \Book{Evolving small-world networks based on
  the modified ba model} in proc. of \Book{2008 International Conference on
  Computer Science and Information Technology} 2008 pp. 143--146.

\bibitem{PhysRevE.95.052303}
\Name{Kryven I.} \REVIEW{Phys. Rev. E}{95}{2017}{052303}.

\bibitem{PhysRevE.89.062814}
\Name{Bianconi G. \and Dorogovtsev S.~N.} \REVIEW{Phys. Rev.
  E}{89}{2014}{062814}.

\bibitem{PhysRevLett.96.040601}
\Name{Dorogovtsev S.~N., Goltsev A.~V. \and Mendes J. F.~F.} \REVIEW{Phys. Rev.
  Lett.}{96}{2006}{040601}.

\bibitem{PhysRevLett.103.058701}
\Name{Newman M. E.~J.} \REVIEW{Phys. Rev. Lett.}{103}{2009}{058701}.

\bibitem{RevModPhys.80.1275}
\Name{Dorogovtsev S.~N., Goltsev A.~V. \and Mendes J. F.~F.} \REVIEW{Rev. Mod.
  Phys.}{80}{2008}{1275}.

\bibitem{boccaletti2006complex}
\Name{Boccaletti S., Latora V., Moreno Y., Chavez M. \and Hwang D.-U.}
  \REVIEW{Phys. Rep.}{424}{2006}{175}.

\bibitem{jeong2000large}
\Name{Jeong H., Tombor B., Albert R., Oltvai Z.~N. \and Barab{\'a}si A.-L.}
  \REVIEW{Nature}{407}{2000}{651–}.

\bibitem{HAJRA200544}
\Name{Hajra K.~B. \and Sen P.} \REVIEW{Physica A}{346}{2005}{44}.

\bibitem{PhysRevE.68.026113}
\Name{Lehmann S., Lautrup B. \and Jackson A.~D.} \REVIEW{Phys. Rev.
  E}{68}{2003}{026113}.

\bibitem{PhysRevLett.87.258701}
\Name{Pastor-Satorras R., V\'azquez A. \and Vespignani A.} \REVIEW{Phys. Rev.
  Lett.}{87}{2001}{258701}.

\bibitem{PhysRevE.66.035103}
\Name{Ebel H., Mielsch L.-I. \and Bornholdt S.} \REVIEW{Phys. Rev.
  E}{66}{2002}{035103(R)}.

\bibitem{bollobas2001random}
\Name{Bollob\'{a}s B.} \Book{Random graphs} 2nd Edition (Cambridge university
  press, Cambridge, England) 2001.

\bibitem{Bollobas80}
\Name{Bollob\'{a}s B.} \REVIEW{Eur. J. Combin.}{1}{1980}{311}.

\bibitem{Steger99}
\Name{Steger A. \and Wormald N.~C.} \REVIEW{Comb. Probab.
  Comput.}{8}{1999}{377}.

\bibitem{erdos1959random}
\Name{Erd\H{o}s P. \and R{\'e}nyi A.} \REVIEW{Publ. Math.-Debr.}{6}{1959}{290}.

\bibitem{freeman1977set}
\Name{Freeman L.~C.} \REVIEW{Sociometry}{40}{1977}{35}.

\bibitem{PhysRevLett.89.208701}
\Name{Newman M. E.~J.} \REVIEW{Phys. Rev. Lett.}{89}{2002}{208701}.

\bibitem{RevModPhys.74.47}
\Name{Albert R. \and Barab\'asi A.-L.} \REVIEW{Rev. Mod. Phys.}{74}{2002}{47}.

\bibitem{1354658}
\Name{Kim M. \and {Medard} M.} \Book{Robustness in large-scale random networks}
  in proc. of \Book{IEEE INFOCOM 2004} Vol.~4 2004 pp. 2364--2373.

\bibitem{PhysRevE.93.062309}
\Name{Nitzan M., Katzav E., K{\"u}hn R. \and Biham O.} \REVIEW{Phys. Rev.
  E}{93}{2016}{062309}.

\bibitem{newmanch2}
\Name{Newman M. E.~J.} \Book{Random graphs as models of networks} (John Wiley
  \& Sons, Ltd) 2005 Ch.~2 pp. 35--68.

\bibitem{PhysRevLett.84.3201}
\Name{Newman M. E.~J., Moore C. \and Watts D.~J.} \REVIEW{Phys. Rev.
  Lett.}{84}{2000}{3201}.

\bibitem{NEWMAN1999341}
\Name{Newman M. E.~J. \and Watts D.~J.} \REVIEW{Phys. Lett. A}{263}{1999}{341}.

\bibitem{stauffer2014introduction}
\Name{Stauffer D. \and Aharony A.} \Book{Introduction to percolation theory}
  2nd Edition (Taylor \& Francis, London) 1994.

\bibitem{Callaway00}
\Name{Callaway D.~S., Newman M. E.~J., Strogatz S.~H. \and Watts D.~J.}
  \REVIEW{Phys. Rev. Lett.}{85}{2000}{5468}.

\bibitem{Cohen00}
\Name{Cohen R., Erez K., {ben-Avraham} D. \and Havlin S.} \REVIEW{Phys. Rev.
  Lett.}{85}{2000}{4626}.

\bibitem{PhysRevLett.85.4104}
\Name{Newman M. E.~J. \and Ziff R.~M.} \REVIEW{Phys. Rev.
  Lett.}{85}{2000}{4104}.

\bibitem{PhysRevE.64.016706}
\Name{Newman M. E.~J. \and Ziff R.~M.} \REVIEW{Phys. Rev. E}{64}{2001}{016706}.

\bibitem{choi2019newman}
\Name{Choi J.-O. \and Yu U.} \REVIEW{J. Comp. Phys.}{386}{2019}{1}.

\bibitem{PhysRevLett.95.098104}
\Name{Santos F.~C. \and Pacheco J.~M.} \REVIEW{Phys. Rev.
  Lett.}{95}{2005}{098104}.

\bibitem{CHEN2007379}
\Name{Chen Y.-S., Lin H. \and Wu C.-X.} \REVIEW{Physica A}{385}{2007}{379}.

\bibitem{szabo.am}
\Name{Hauert C. \and Szab\'{o} G.} \REVIEW{Am. J. Phys.}{73}{2005}{405}.

\bibitem{PhysRevE.79.016106}
\Name{Szab\'o G. \and Szolnoki A.} \REVIEW{Phys. Rev. E}{79}{2009}{016106}.

\bibitem{PhysRevE.73.067103}
\Name{Vukov J., Szab\'o G. \and Szolnoki A.} \REVIEW{Phys. Rev.
  E}{73}{2006}{067103}.

\bibitem{PhysRevE.72.047107}
\Name{Szab\'o G., Vukov J. \and Szolnoki A.} \REVIEW{Phys. Rev.
  E}{72}{2005}{047107}.

\bibitem{PhysRevLett.98.108103}
\Name{G\'omez-Garde\~nes J., Campillo M., Flor\'{\i}a L.~M. \and Moreno Y.}
  \REVIEW{Phys. Rev. Lett.}{98}{2007}{108103}.

\bibitem{Xia_2015}
\Name{Xia C.-Y., Meloni S., Perc M. \and Moreno Y.} \REVIEW{{EPL} (Europhysics
  Letters)}{109}{2015}{58002}.

\bibitem{chen2016evolution}
\Name{Chen M.-h., Wang L., Sun S.-w., Wang J. \and Xia C.-y.} \REVIEW{Phys.
  Lett. A}{380}{2016}{40}.

\bibitem{chen2015impact}
\Name{Chen M.-h., Wang L., Wang J., Sun S.-w. \and Xia C.-y.} \REVIEW{Appl.
  Math. Comput.}{251}{2015}{192}.

\bibitem{perc2017statistical}
\Name{Perc M., Jordan J.~J., Rand D.~G., Wang Z., Boccaletti S. \and Szolnoki
  A.} \REVIEW{Phys. Rep.}{687}{2017}{1}.

\bibitem{li2018punishment}
\Name{Li X., Jusup M., Wang Z., Li H., Shi L., Podobnik B., Stanley H.~E.,
  Havlin S. \and Boccaletti S.} \REVIEW{Proc. Natl. Acad. Sci.
  U.S.A.}{115}{2018}{30}.

\bibitem{wang2015evolutionary}
\Name{Wang Z., Wang L., Szolnoki A. \and Perc M.} \REVIEW{Eur. Phys. J.
  B}{88}{2015}{124}.

\bibitem{li2019effect}
\Name{Li Z., Jia D., Guo H., Geng Y., Shen C., Wang Z. \and Li X.}
  \REVIEW{Appl. Math. Comput.}{351}{2019}{162}.

\bibitem{nowak1992evolutionary}
\Name{Nowak M.~A. \and May R.~M.} \REVIEW{Nature}{359}{1992}{826}.

\end{thebibliography}
\bibliographystyle{eplbib}
\end{document}